\documentclass[english]{elsarticle}
\usepackage[T1]{fontenc}
\usepackage[latin9]{inputenc}
\usepackage{rotfloat}
\usepackage{amsmath}
\usepackage{amssymb}
\usepackage{graphicx}
\usepackage{color}
\usepackage{cite}

\usepackage[linesnumbered,lined,boxed,commentsnumbered]{algorithm2e}
\usepackage{subfig}

\journal{Journal of Knowledge-Based Systems}
\makeatletter

\newtheorem{thm}{Theorem}

\newdefinition{rmk}{Remark}
\newproof{pf}{Proof}
\newproof{pot}{Proof of Theorem \ref{thm2}}

\@ifundefined{showcaptionsetup}{}{%
 \PassOptionsToPackage{caption=false}{subfig}}

\begin{document}
\begin{frontmatter}
\title{Two Evidential Data Based Models for Influence Maximization in Twitter}

\author[ISG,IRISA,CERT]{Siwar Jendoubi\corref{mycorrespondingauthor}}
\cortext[mycorrespondingauthor]{Corresponding author}
\ead{jendoubi.siwar@yahoo.fr}
\author[IRISA]{Arnaud Martin}
\author[IRISA]{Ludovic Liétard}
\author[CERT]{Hend Ben Hadji}
\author[IHEC]{Boutheina Ben Yaghlane}

\address[ISG]{Université de Tunis, ISG Tunis, LARODEC}
\address[IRISA]{Université de Rennes I, IRISA}
\address[IHEC]{Université de Carthage, IHEC Carthage, LARODEC}
\address[CERT]{Centre d'Etude et de Recherche des Télécommunications}

\begin{abstract}
Influence maximization is the problem of selecting a set of influential users in the social network. Those users could adopt the product and trigger a large cascade of adoptions through the ``word of mouth''
effect. In this paper, we propose two evidential influence maximization models for Twitter social network. The proposed approach uses the theory of belief functions to estimate users influence. Furthermore, the proposed influence estimation measure fuses many influence aspects in Twitter, like the importance of the user in the network structure and the popularity of user's tweets (messages). In our experiments, we compare the proposed solutions to existing ones and we show the performance of our models.
\end{abstract}


\end{frontmatter}


\section{Introduction}
Social influence is defined by ``changes in an individual's thoughts, feelings, attitudes, or behaviors that result from interaction with another individual or a group'' \citep{Rashotte07}. Identifying influencers in online social networks has received great attention from researchers in many research fields like sociology, marketing, psychology, computer science, etc. For example, marketers look for influencers to promote their marketing campaign. In fact, influencers are able to make the product propaganda goes viral through the social network, therefore, we are on the brink of defining the problem of influence maximization.

The problem of influence maximization has been widely studied since its introduction \citep{Domingos01,Kempe03,Goyal10,Goyal12}. Its purpose is to select a set of $k$ influential users in the social network that could adopt the product and trigger a large cascade of adoptions through the ``word of mouth'' effect. Consider the following example; a startup company produces a new product and wants to market it and it has a small budget for that purpose. A good solution may be to get profit from online social networks (OSN) through the ``word of mouth'' effect. Hence, it can select a small number of initial users, it encourages them to adopt the product, for example by giving them gifts or discounts. Selected users start spreading what through the network to influence their friends and their friends influence their friend's friends until achieving reaching instead as large individuals as possible. To that end a question arises: how do the startup company select the initial set of users to maximize the awareness of the product. The main goal of the influence maximization is to find solutions to that problem.

The authors in \citep{Domingos01} were the firsts to introduce the problem of identifying influencers for a marketing campaign as a learning problem. They modeled the customer's network value, \textit{i.e.} ``the expected profit from sales to other customers he may influence to buy, the customers those may influence, and so on recursively'' \citep{Domingos01} , also they considered the market as a social network of customers. Motivated by the work of \citep{Domingos01}, Kempe \textit{et al.} \citep{Kempe03} formulated the problem as an optimization problem which is proven to be NP-Hard. They assumed having  the social network and the influence probabilities extent to which each individual influence one another. Their issue is to find/choose a set of influential individuals that maximize the spread of the marketing message within the network.

Real world is full of imprecision and uncertainty and this fact will necessarily reflect on OSN data. The imprecision of the information is characterized by its content. In fact, it is related to the information or to the source. It measures a quality issue of the knowledge. The uncertainty of the information characterizes the degree of its conformity to the reality. Therefore, an uncertain information describes a partial knowledge of the reality.  In fact, social interactions can not be always precise and certain, also, OSNs allow only limited access to their data which generates more imprecision and uncertainty for the social network analysis fields. 
The uncertainty is due to the partial knowledge we have about the user. For example, we do not have all the user's tweets or all his relations in the network. It leads to imprecise measurements. For example, it is not possible to obtain a precise information about the user's opinion because we do not have all his tweets. Then, if we ignore this imperfection of the data, we may be confronted to obtain erroneous analysis results. In such a situation, the theory of belief functions \citep{Dempster67a,Shafer76} has been widely applied, and it is able to well characterize the uncertain (ignorant) information and reduce the errors. We find it used, for example, in some related research fields like pattern clustering \citep{Denoeux2016,Liu2015} and classification \citep{Liu2016}. Furthermore, this theory was used for analyzing social networks \citep{Wei13,Gao13,Kim2013,Jendoubi14a,Jendoubi2015,Zhouab2015}.

In this paper, we propose two evidential influence maximization models for Twitter social network. We use the theory of belief functions to estimate the influence taking into account data imperfection. The proposed approach benefits from the performance of the mathematical framework of belief functions especially the fusion of the information. In fact, our influence measure consider many influence aspects, like the importance of the user in the network structure and the popularity of user's tweets (messages), then, this theory manages all these aspects, fuses them and deals with uncertainty and imprecision that characterize social network data.
The resulting measure is obviously more refined and more precise than taking each influence aspect separately. 

Our work achieves the following contributions: 
\begin{enumerate}
	\item We propose a new evidential influence measure for the Twitter social network%
\footnote{www.twitter.com}, the proposed measure considers many influence aspects which makes
it more refined and precise than existing measures that are proposed
for Twitter. 
	\item We use the theory of belief functions to combine influence indicators like the number of followers, the number of retweets and the number of mentions, that are generally considered separately  while the analysis of the influence on Twitter. 
	\item  We introduced a new influence aspect which is summarized {by} the fact that ``I am
more influencer if I am connected to influencer users'' and we considered it in our measure.
	\item We maximized the influence based on the proposed influence measure.
	\item We show that influence maximization under our model is NP-Hard. Nevertheless, we show that the function defining the influence propagation is monotone and sub-modular. Consequently, we develop a greedy based algorithm that guarantees a good approximation to the optimal solution. 
	\item We conduct our experiments on real word data set that we collected from Twitter.

\end{enumerate}
 
The remainder of this paper is organized as {follows}: section 2 introduces
some basic concepts of the theory of belief functions, section 3 discusses
some related works, section 4 presents the proposed evidential influence
maximization model and section 5 provides results from our experiments.

\section{Background: Theory of belief functions}

In this section, we introduce some basic concepts of the theory of
belief functions that were used in the proposed approach. Dempster
\citep{Dempster67a} proposed the \textit{Upper and Lower
probabilities} that are considered as the first ancestor of the evidence
theory, also called Dempster-Shafer theory or belief functions theory.
Then \citep{Shafer76} introduced the \textit{Mathematical theory
of evidence} and defined the basic mathematical framework of the evidence
theory, often called \textit{Shafer model}. The main goal of the Dempster-Shafer
theory is to achieve more precise, reliable and coherent information.

Let $\Omega=\left\{ s_{1},s_{2},...,s_{n}\right\} $ be the \textit{frame
of discernment}. The basic belief assignment (BBA),
$m^{\Omega}$, represents the agent belief on $\Omega$, it is defined as: 
\begin{eqnarray}
2^{\Omega} & \rightarrow & \left[0,1\right]\nonumber \\
A & \mapsto & m^{\Omega}\left(A\right)
\end{eqnarray}
 where $2^{\Omega}=\left\{ \emptyset,\left\{ s_{1}\right\} ,\left\{ s_{2}\right\} ,\left\{ s_{1},s_{2}\right\} ,...,\left\{ s_{1},s_{2},...,s_{n}\right\} \right\} $
is the \textit{power set} (set of all subsets) of $\Omega$. $m^{\Omega}\left(A\right)$
is the mass value assigned to $A\subseteq\Omega$. The mass function
$m^{\Omega}$ must respect: 
\begin{equation}
\sum_{A\subseteq\Omega}m^{\Omega}\left(A\right)=1\label{eq:mass}
\end{equation}
 In the case where we have $m^{\Omega}(A)>0$, $A$ is called \textit{focal element} of $m^{\Omega}$. 
The mass value given to the set $\Omega$,  $m^{\Omega}(\Omega)$,  
  is the mass that cannot be given to its subsets and it is called total ignorance. When we compare a BBA distribution to a probability distribution, we notice that the BBA allows a subset of $\Omega$
  to be a focal element when we have some doubt about the decision, while the probability theory forces the equiprobability in such a case.

Combination rules in the evidence theory are the main tool that can
be used for information fusion. \textit{Dempster's rule}
was the first defined in this theory \citep{Dempster67a}, it is used
to fuse two distinct BBAs, defined on $\Omega$, that come from two
different sources describing the same event. It is defined as: 
\begin{equation}
m_{1\oplus2}^{\Omega}\left(A\right)=\begin{cases}
\frac{\displaystyle\sum_{B\cap C=A}m_{1}^{\Omega}\left(B\right)m_{2}^{\Omega}\left(C\right)}{1-\displaystyle\sum_{B\cap C=\emptyset}m_{1}^{\Omega}\left(B\right)m_{2}^{\Omega}\left(C\right)}, & A\subseteq\Omega\setminus\left\{ \emptyset\right\} \\
0 & \textrm{if}\, A=\emptyset
\end{cases}\label{eq:dempster rule}
\end{equation}

To make a decision with the belief functions framework, we can use
the pignistic transformation \citep{Smets05b} to get a probability
distribution from a BBA distribution. Then, for each element in the
frame of discernment $\Omega$, we compute its pignistic probability
as follows:

\begin{equation}
\textrm{BetP}^{\Omega}\left(s_{i}\right)=\sum_{A\in2^{\Omega},s_{i}\in A}\frac{m^{\Omega}\left(A\right)}{\mid A\mid\left(1-m\left(\emptyset\right)\right)},\,\, s_{i}\in\Omega\label{eq:BetP}
\end{equation}

\section{Related work}

Influence maximization (IM) is the problem of finding
a set of $k$ seed nodes that are able to influence the maximum number
of nodes in the social network. In the literature, we find many solutions
for the IM problem. In this section, we present an overview of the
state of the art. First, we introduce the influence maximization basic
models that {use} a diffusion model, then we present data based models.
After, we talk about influence in Twitter and finally we  present
measures that {use} the theory of belief functions to estimate influence.

\subsection{Diffusion model based influence maximization}

Given a social network $G=\left(V,E\right)$, $V$ is a set of vertices,
$E$ is a set of edges and a diffusion model $M$, the influence maximization
(IM) problem is to select a set $S$ of $k$ influential users (called
seed set) that maximizes the awareness of the ``product'' over the
social network $G$ \citep{Kempe03}. In other words, it is the problem
of choosing $S$ seed nodes that {maximize} the expected number of
influenced nodes, $\sigma_{M}\left(S\right)$, that will adopt the
``product''. Maximizing $\sigma_{M}\left(S\right)$ is a NP-Hard
problem. Authors in \citep{Kempe03} prove that $\sigma_{M}\left(S\right)$ is
monotone, \textit{i.e.} 
\begin{equation}
\sigma_{M}\left(S\right)\leq\sigma_{M}\left(T\right)
\end{equation}
whenever $S\subseteq T\subseteq V$ and sub-modular, \textit{i.e.}
\begin{equation}
\sigma_{M}\left(S\cup\left\{ x\right\} \right)-\sigma_{M}\left(S\right)\geq\sigma_{M}\left(T\cup\left\{ x\right\} \right)-\sigma_{M}\left(T\right)
\end{equation}
whenever $S\subseteq T\subseteq V$ and $x\in V$, hence, they used the
greedy algorithm with the Monte Carlo simulation to extract the seed
set. To estimate $\sigma_{M}\left(S\right)$, \citep{Kempe03} propose
two propagation simulation models which are the\textit{ Linear Threshold
Model (LTM) }\citep{Granovetter78} and the \textit{Independent Cascade
Model (ICM)} \citep{Goldenberg01}. In these models we suppose that
we have a social graph $G=\left(V,E\right)$, a vertex $v$ is said
to be \textit{active} if it received the information and accepted
it. It is said to be \textit{inactive} if it did not receive the information
or rejected it. An inactive node can become active. In the LTM model
we associate a \textit{weight} to each edge $\omega\left(u,v\right)$
and a \textit{threshold} $\theta_{u}$ to each vertex. A vertex $u$
will be activated if the total weight, between it and its activated
neighbors, is at least $\theta_{u}$, \textit{i.e.} 
\begin{equation}
\sum_{v}\omega\left(u,v\right)\geq\theta_{u}
\end{equation}
The threshold $\theta_{u}$ is a random uniform variable chosen from
$\left[0,1\right]$, it ``intuitively {represents} the different latent
tendencies of nodes to adopt the innovation when their neighbors do''
\citep{Kempe03}. In the ICM model each newly activated node is given
only one chance to activate its inactive neighbors. For instance,
at the step $t$, a newly activated node $u$ will try to activate
its inactive neighbor $v$, the success probability of $u$ to activate
$v$ is given by $p\left(u,v\right)$ (parameter of the system). A
special case of ICM is \textit{Weighted Cascade (WC)} where 
\begin{equation}
p\left(u,v\right)=\frac{1}{D_{u}}
\end{equation}
 such that $D_{u}$ is the overall degree of the vertex $u$. 

In the literature, many works were conducted to improve the running
time when considering ICM and LTM. The work of \citep{Leskovec07b} introduced
the Cost Effective Lazy Forward (CELF) algorithm. It exploited the
sub-modularity property of the function to be maximized and proved
to be 700 times faster than the solution of \citep{Kempe03}. Authors in \citep{Mathioudakis11}
introduced SPIN (Sparcification of influence network). It is an instance
of the ICM, that reduces the complexity by network simplification.
It starts by selecting the $k$ edges that are most likely to explain
the propagation, then it applies the greedy algorithm in order to
select arcs that increase the likelihood. After network simplification,
it applies the ICM algorithm to detect users that maximize influence. The work of \citep{Kim2014} presented the 
``Continuously activated and Time-restricted IC (CT-IC) model'' that generalizes the ICM. CT-IC model gives {to} each active node many chances to activate its neighbors and these chances are processed until a given time.
Other works tried to consider other parameters to improve the quality
of the selected seed nodes. Among these parameters we find the topic
\citep{Aslay14,Barbieri12}, trust \citep{Ahmed2013,Baghmolaei},
time \citep{Liu12}, etc.

\subsection{Data based influence maximization}

In the literature, we find that influence probabilities are either
uniform \textit{i.e.} $p\left(u,v\right)=0.01$, or selected uniformly
at random from the set \\
$\left\{ 0.01,\,0.001,\,0.0001\right\} $ or computed as in the weighted
cascade \textit{i.e.} $p\left(u,v\right)=\frac{1}{D_{u}}$ ($D_{u}$
is the overall degree of the vertex $u$). The work of \citep{Goyal10}
{introduces} many data based models to learn influence probabilities.
In their paper they consider the static case, the continuous time
case and the discrete time case, for more details the reader can refer
to \citep{Goyal10}. 

The credit distribution (CD) \citep{Goyal12} is, also, a data based
approach that investigates past propagation to detect users of influence.
It uses past propagation actions to associate an influence credit
to each user in the network. The influence spread is defined as the
total influence credit given to a set of users $S$ from the {whole}
network. The idea behind this algorithm is that; when an action $a$
propagates from a user $u$ to a user $v$, a direct influence credit,
$\gamma\left(u,v\right)\left(a\right)$, is given to $u$. Also, a
credit amount is given to predecessors of $u$ in the propagation
graph. The first step of the credit distribution algorithm consists
on scanning the action log $L$ (a data structure that is defined
as the set of tuples $\left(User,\, Action,\, Time\right)$ such that
$\left(u,a,t\right)\in L$ means that the user $u$ performed the
action $a$ at time $t$) to compute the total credit given to $u$
for influencing its neighbor $v$ for the action $a$, $\Gamma\left(u,v\right)\left(a\right)$.
$S$ the set of seed nodes is initialized to $\emptyset$. In the
second step, the algorithm runs up the CELF algorithm to select the
node with the maximum marginal gain and so on until getting all needed
seed nodes. For more details the reader can refer to \citep{Goyal12}.

As the work of \citep{Goyal12}, our work is data based. However,
our approaches differ from it in the following ways. First, we propose
an influence measure that {considers} many influence aspects like the
structure of the network, the influence of the user's friends, the
user's popularity etc. Second, we use the theory of belief functions
to combine all pieces of information from each influence aspect in
order to model uncertainty and imprecision and to manage the conflict
between {the pieces of information}.

\subsection{Influence in Twitter}

Actually, Twitter is one of the most popular micro-blogging {services}.
It allows its users to follow updates from each other via the ``follow''
relationship. For example, let $A$ and $B$ be two Twitter users,
then, if $A$ is interested by updates from $B$, {$A$} can simply ``follow''
{it} and {$A$} will {receive} all the messages (called tweets) from $B$
in {its} actuality time-line, also, Twitter users can have access to
public tweets that {appear} in a public time-line. The follow relationship
can either be reciprocated or one way. Twitter enables its users to
send and read short 140-character messages called ``tweets''. Besides,
{Twitter users} can spread tweets from others and share them with {their} own followers
using the ``retweet'' mechanism. Furthermore, users are able to
send tweets directly to other users by mentioning their username prefixed
with an ``@'' sign. 

In the literature, influence in Twitter was widely studied. In \citep{Cha10}
the authors present an in-depth comparison of three influence measures
which are indegree (follow), retweets and mentions. Authors in \citep{Brown2011}
measure the user influence in Twitter using the K-Shell decomposition
algorithm that takes as input the followership network and gives as
a result an influence value for each user, also, they modify the basic
algorithm to assign to each user a logarithmic K-Shell influence value.
The work of \citep{Sung2013} {proposes} InterRank measure that improves
the PageRank measure by considering not only the follower relationship
of the network but also the topical similarity between Twitter users.
In \citep{BenJabeur2012} the authors define Twitter influencers as
``active actors who have the ability to spread information and inspire
other people in the network'' and they propose InfRank algorithm
that identifies influencers according to their retweet activity. In
\citep{Dubois2014} authors compare six influence metrics (like indegree,
eigenvector centrality and clustering coefficient) that are commonly
used to identify influential users in Twitter. The work of \citep{Rudat2015}
{studies} the influence of the ``information value'' of the tweet (content
criteria) and the agent awareness (context criteria) on the retweeting
decision.

To sum up, in this section we presented works that searched to estimate
influence in Twitter. All these works used only one {criterion} in each
proposed measure, \textit{i.e.} \citep{Brown2011} used the follow
relationship, \citep{BenJabeur2012} used the retweet relationship.
Some of these works tried to compare influence measures separately
like the work of \citep{Cha10} and \citep{Dubois2014}. To the best
of our knowledge, our work is the first that introduces an influence
measure that fuses many influence {aspects} in Twitter and we used the
proposed measure to maximize influence.

\subsection{Influence and evidence theory}

In the literature, we find two works \citep{Wei13,Gao13} that talk
about identifying influence nodes under the framework of the evidence
theory. The work of \citep{Wei13} defines an evidential centrality
(EVC) measure that works in a weighted network, \textit{i.e.} network
centrality is a measure that searches to identify important nodes in
the network according to one or more criteria. They {define} two BBAs
distributions on the frame $\left\{ high,\, low\right\} $, \textit{i.e.}
high influence and low influence, for each node in the network. The
first BBA is used to measure the degree centrality and the second
one is used to measure the strength centrality of the node. Finally,
the proposed centrality measure is the result of the combination of
these BBAs. 

Authors in \citep{Gao13} {propose} {another} centrality measure that
avoids some drawbacks of the measure of \citep{Wei13}. In fact, they
{modify} the EVC measure according to the actual degree of the node
instead of following the uniform distribution, also, they {extend}
the semi-local centrality measure \citep{Chen12} to be used with
weighted networks. Their centrality measure is the result of the combination
of the modified EVC and the modified semi-local centrality measure.
Authors in \citep{Gao13} and in \citep{Wei13} use the same frame
of discernment, their measures are structure based and they choose
the influential nodes to be top-1 ranked nodes according to the proposed
centrality measure. 

In this paper, we use the BBA estimation mechanism of \citep{Wei13,Gao13} in our approach.
Nevertheless, our influence measure fuses more influence parameters.

\section{Proposed evidential influence maximization models}

In this paper, we propose two new models to maximize {the} influence {on} twitter social network. We use the theory of belief functions to overcome
the problem of data imperfection. {In fact,} twitter API does not
allow the access to all Twitter data, in fact there {are} a limited number
of data requests by hour which causes the imperfection of the data:
uncertainty, imprecision, lack of data, \textit{etc}, and to fuse
many influence aspects in Twitter that were studied separately in several
works in the literature like the work of \citep{Cha10} and \citep{Dubois2014}.
Furthermore, we assume that an influencer on Twitter has to
be: active by tweeting frequently, followed by several users in the
network that are interested by his actuality, frequently mentioned
in other users' tweets and his tweets are retweeted many times. In this
section we present two evidential influence maximization models that
consider our assumption while measuring user's influence.

\subsection{Weights computation}

\begin{figure}
\begin{centering}
\includegraphics[scale=0.75]{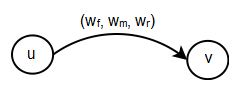}
\par\end{centering}

\caption{Weight vector between $u$ and $v$.\label{fig:Weight-vector}}

\end{figure}

In Twitter social network there are three possible relations: the first
one is explicit which is the follow relation, {\textit{i.e.} the follow relation is created when a given user follows another user}, the second and the third
relations are implicit which are the mention and the retweet, {\textit{i.e.} we obtain these implicit relations when we collect the user's tweets, then when a user mentions or retweets another user we create a new implicit link or we update an existing one}. Another
property of Twitter, is that between two {given} users $u$ and $v$ we can
have a follow, a mention and/or a retweet relation. {We assign to each link $\left(u,v\right)$ a vector of three weights, \textit{i.e.} follow weight, mention weight and retweet weight,} that has the form $\left(w_{f},w_{m},w_{r}\right)$
as shown in Figure (\ref{fig:Weight-vector}). The follow weight $w_{f}$
measures the strength of the followership between $u$ and $v$,\textit{
i.e.} when the direct followership relation is broken, $w_{f}$ measures
the fact that $u$ still receives $v$'s tweets via intermediary users
between them. The mention weight $w_{m}$ weights information exchange
between users $u$ and $v$, indeed, when $u$ mentions $v$ in a
tweet then this second ($v$) will receive directly the message in
his notification tab. This behavior emphasizes direct communication
between twitter users. The retweet weight $w_{r}$ represents the
information diffusion and influence weight between users, in fact,
more $v$ retweets from $u$ more it is influenced by $u$ \citep{BenJabeur2012}.

\begin{figure}
	\centering
		\includegraphics[width=0.50\textwidth]{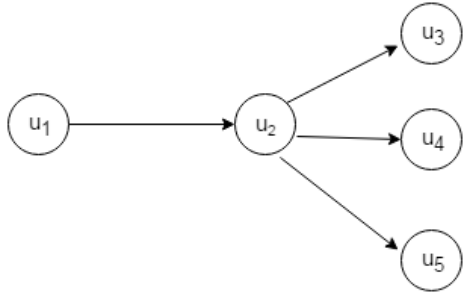}
	\caption{Follow weight example}
	\label{fig:uvlink}
\end{figure}

{Let $G(V,E)$ be the social network where $V$ is the set of nodes and $E$ is the set of links.}
 Let $S_{u} {\subseteq V}$ be the set of immediate successor of $u {\in V}$, $P_{{v}} {\subseteq V}$ the
set of immediate predecessor of {$v \in V$}, $T_{u}$ the set of tweets of
$u$, $R_{u}\left(v\right)$ the set of tweets of $u$ that were retweeted
by $v\in V$, $M_{u}\left(v\right)$ the set of tweets of $u$ in
which $v$ was mentioned and $M_{u}$ the set of tweets in which $u$
mentions any user in the network except himself. {Ben Jabeur \textit{et al.} \citep{BenJabeur2012}} define weights
of the follow relation, mention relation and retweet relation respectively
as {follows} with $v\in S_{u}$:

\begin{eqnarray}
w_{f}\left(u,v\right) & = & \frac{|S_{u}\cap P_{v}|+1}{|S_{u}|}\label{eq:Follow-weight}\\
w_{m}\left(u,v\right) & = & \frac{|M_{u}\left(v\right)|}{|M_{u}|}\\
w_{r}\left(u,v\right) & = & \frac{|R_{u}\left(v\right)|}{|T_{u}|}\label{eq:reweetweight}
\end{eqnarray}
These measures propose to estimate the link weights at a local level,
\textit{i.e.} relatively to the source of the link. We noticed that
Ben Jabeur \textit{et al.} weights are not suitable to our case. Indeed,
in the case {where} $u$, the source of the link, has few successors,
\textit{i.e. }small $S_{u}$, then its out links will get high follow
weights and the same goes for mention and retweet weights. This fact
causes erroneous results. In fact, we may be confronted to obtain
users that have high influence value but they are not active \textit{i.e.
} with small $|S_{u}|$, small $|M_{u}|$ and small $|T_{u}|$. 
Let's take the example in Figure \ref{fig:uvlink}. If we use the equation \ref{eq:Follow-weight} to estimate 
$w_{f}\left(u_{1},u_{2}\right)$, we will obtain $w_{f}\left(u_{1},u_{2}\right)=1$. The value of $w_{f}\left(u_{1},u_{2}\right)$ means that the relationship between $u$ and $v$ is very strong, \textit{i.e.} if the direct link between them is broken $v$ still receive news from $u$, which is not the case.

To remedy this problem, we modify the definitions of \citep{BenJabeur2012} and we estimate the weights with respect to the whole network as {follows}:

\begin{eqnarray}
w_{f}\left(u,v\right) & = & \frac{|S_{u}\cap {P_{v}}|+1}{|S_{max}|}\\
w_{m}\left(u,v\right) & = & \frac{|{M_{v}\left(u\right)}|}{|M_{max}|}\\
w_{r}\left(u,v\right) & = & \frac{|R_{u}\left(v\right)|}{|T_{max}|}
\end{eqnarray}
such that $S_{max}=\max_{u\in V}S_{u}$, $M_{max}=\max_{u\in V}M_{u}$, $T_{max}=\max_{u\in V}T_{u}$ {and  $M_{v}\left(u\right)$ is the set of tweets of $v$ in which $u$ was mentioned.}

{In the next step}, we need to estimate {the} weights
in the node level. {In fact, we compute the three weights for each node in the network, \textit{i.e.} we compute a follow weight, a retweet weight and a mention weight.} Thus, for each node in the network we sum its out links weights as:

\begin{equation}
w_{x}\left(u\right)=\sum_{v\in V}w_{x}\left(u,v\right)
\end{equation}
{where $w_{x}\left(u\right)\in\left\{ w_{f}\left(u\right),\, w_{r}\left(u\right),\, w_{m}\left(u\right)\right\}$ 
  and \\
	$w_{x}\left(u,v\right)\in\left\{ w_{f}\left(u,v\right),\, w_{r}\left(u,v\right),\, w_{m}\left(u,v\right)\right\}$.}
	
\begin{figure}
\begin{centering}
\includegraphics[scale=0.65]{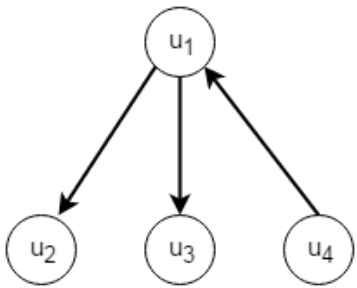}
\par\end{centering}

\caption{Network example \label{fig:Weights-example}}

\end{figure}

	{Let's take the network example given in Figure \ref{fig:Weights-example},
in this example, we have a social network of four users related to
each other by three links. Suppose that after applying the process
of link weights estimation described above for each link, we obtain
weights given in Table \ref{tab:Links-weights}. To compute each node
weights, we sum up its outlinks weights, then the follow weight of
the node $u_{1}$ is $w_{f}\left(u_{1}\right)=w_{f}\left(u_{1},u_{2}\right)+w_{f}\left(u_{1},u_{3}\right)=0.3+0.4=0.7$. Nodes weights are given in Table \ref{tab:Node-weights}.}

\begin{table}
\begin{centering}
\subfloat[Links weights\label{tab:Links-weights}]{\begin{centering}
\begin{tabular}{|c|c|c|c|}
\hline 
Link & $w_{f}$ & $w_{m}$ & $w_{r}$\tabularnewline
\hline 
$\left(u_{1},u_{2}\right)$ & 0.3 & 0.4 & 0.2\tabularnewline
\hline 
$\left(u_{1},u_{3}\right)$ & 0.4 & 0.3 & 0.1\tabularnewline
\hline 
$\left(u_{4},u_{1}\right)$ & 0.5 & 0.4 & 0.3\tabularnewline
\hline 
\end{tabular}
\par\end{centering}

} ~~\subfloat[Nodes weights\label{tab:Node-weights}]{\begin{centering}
\begin{tabular}{|c|c|c|c|}
\hline 
Node & $w_{f}$ & $w_{m}$ & $w_{r}$\tabularnewline
\hline 
$u_{1}$ & 0.7 & 0.7 & 0.3\tabularnewline
\hline 
$u_{2}$ & 0 & 0 & 0\tabularnewline
\hline 
$u_{3}$ & 0 & 0 & 0\tabularnewline
\hline 
$u_{4}$ & 0.5 & 0.4 & 0.3\tabularnewline
\hline 
\end{tabular}
\par\end{centering}

}
\par\end{centering}

\caption{Links and nodes weights\label{tab:Links-and-nodes}}

\end{table}

We attract the reader's attention to the fact that we use the sum
function to aggregate users weights for its simplicity, but it is possible
to use other aggregation functions like the mean for example. In the
next section, we focus on the influence estimation and we present
step by step our method for estimating influence.

\subsection{Influence estimation}

In this section, we present our method to estimate the influence. {We introduce a new influence measure for Twitter users, the novelty of this measure is that it is an evidential measure and it contracts many influence aspects.}
Let \\ $\Omega=\left\{ I,\, P\right\} $ be our frame of discernment:
\textit{I} models the user's influence and \textit{P} the user's passivity,
a user cannot be influencer and passive at the same time, and let
$G=\left(V,\, E,\, W\right)$ be a directed graph where {$V$ is the set of nodes, \textit{i.e.}} $v\in V,\, u\in V$
are nodes in $G$, {$E$ is the set of links, \textit{i.e.}} $\left(u,v\right)\in E$ is the {link} that has $u$ as {a} source and $v$ as {a} destination and {$W$ is the set of weights vectors, \textit{i.e.}} $\left(w_{f}\left(u,v\right),\, w_{m}\left(u,v\right),\, w_{r}\left(u,v\right)\right)\in W$
is the weight vector associated to $\left(u,v\right)$. The influence
estimation process contains three basic steps: in the first step,
we estimate a BBA distribution for each node in the network, this
BBA summarizes many influence aspects that are related to the node.
In the second step, for each node {$u$} we use its estimated BBA (the result
of step one) to update its in-links weights, \textit{i.e.} links having $u$ as destination. In the last step, we use the updated weights to estimate a BBA distribution that contracted
many influence {aspects}.

\paragraph{Step 1: Node level}

Let $N_{min_{x}}=\min_{u\in V}w_{x}\left(u\right)$ and $N_{max_{x}}=\max_{u\in V}w_{x}\left(u\right)$.
Then, for each node in the network, we estimate a mass distribution
for each variable, \textit{i.e.} Follow, Mention and Retweet, using their weights.
For each \\ $u\in V$, and for each weight $w_{x}\left(u\right)\in\left\{ w_{f}\left(u\right),w_{r}\left(u\right),w_{m}\left(u\right)\right\} $,
we estimate a mass distribution as follows \citep{Wei13,Gao13}: 

\begin{eqnarray}
m_{x_{{\left(u\right)}}}^{\Omega}\left(I\right) & = & \frac{w_{x}\left(u\right)-N_{min_{x}}}{\gamma_{x}}\\
m_{x_{{\left(u\right)}}}^{\Omega}\left(P\right) & = & \frac{N_{max_{x}}-w_{x}\left(u\right)}{\gamma_{x}}\\
m_{x_{{\left(u\right)}}}^{\Omega}\left(\left\{ I,P\right\} \right) & = & 1-\left(m_{x_{{\left(u\right)}}}^{\Omega}\left(I\right)+m_{x_{{\left(u\right)}}}^{\Omega}\left(P\right)\right)
\end{eqnarray}
where $\gamma_{x}=N_{max_{x}}-N_{min_{x}}+\alpha$, $\alpha\in\left[0,1\right]$. {The mass value given to the set $\Omega=\left\{ I,P\right\}$  is the total ignorance mass value that cannot be given to singletons. In fact this mass models the uncertainty and the imprecision.}
At the end of this step, we have three BBA distributions defined on
$\Omega$, \textit{i.e.} follow BBA, mention BBA and retweet BBA,
for each node in the network. Then, we combine all these BBAs using
the Dempster's rule of combination (equation (\ref{eq:dempster rule})),
\textit{i.e.} \\ $m^{\Omega}_{{\left(u\right)}}=\left(m_{f_{{\left(u\right)}}}^{\Omega}\oplus m_{r_{{\left(u\right)}}}^{\Omega}\right)\oplus m_{m_{{\left(u\right)}}}^{\Omega}$.
After this step, we apply the pignistic transformation on the resulting
combined BBA $m^{\Omega}_{{\left(u\right)}}$ (equation (\ref{eq:dempster rule})).
We obtain a pignistic probability distribution $BetP^{\Omega}_{{\left(u\right)}}$
(equation (\ref{eq:BetP})). At the end of this step, we have a probability
value for each node that reflects many influence aspects such that:
\begin{enumerate}
\item The importance of the user in the network structure. Indeed, the number
of user's followers in the Twitter network reflects his structural
importance.
\item The popularity of user's tweets that we measure using the number of
times where user's tweets are retweeted. 
\item The popularity of the user that can be measured by the number of times
the user was mentioned in other user's tweet. In fact, we assume that
more the user is mentioned more he is popular in the network.
\end{enumerate}

\paragraph{Step 2: Updating weights\label{par:Step-2:-Updating}}

The main {contribution} of this second step is to take into account the following assumption: ``\textit{I am more influencer if I am connected to influencer users}''. It means that when a given user is connected to other influencers, his personal influence increases. To consider this assumption, we update weights vector of each link in the network using the estimated pignistic probability distributions defined on the link destination node:

\begin{equation}
w_{x}^{'}\left(u,v\right)=w_{x}\left(u,v\right).BetP_{{\left(v\right)}}^{\Omega}\left(I\right)
\end{equation}
where $w_{x}\left(u,v\right)\in\left\{ w_{f}\left(u,v\right),\, w_{r}\left(u,v\right),\, w_{m}\left(u,v\right)\right\} $
and \\
$w_{x}^{'}\left(u,v\right)\in\left\{ w_{f}^{'}\left(u,v\right),\, w_{r}^{'}\left(u,v\right),\, w_{m}^{'}\left(u,v\right)\right\} $
is the vector of updated link weights. In this equation, we ponder
the weight value given to the influence link between $u$ and $v$
by the influence pignistic probability of the destination node $v$,
$BetP^{\Omega}_{{\left(v\right)}}\left(I\right)$. Using this step, the node $v$ propagates its influence to its in-neighbors, \textit{i.e.} neighbors having $v$ as destination. Then, if the
influence of $v$ is high, the weights of its in-links will maintain
a high value from their original amount and if the influence of $v$
is low, the weights of its in-links will maintain only a low value from their
influence before the updating. Therefore, if a user $u$ is connected to
many influencer users, then, his own influence will be consolidated
using the proposed equation.

\paragraph{Step 3: Link level}

In this step, we estimate a mass distribution for each weight value and for each link $\left(u,v\right)\in E$ as {follows}:

\begin{eqnarray}
m_{x_{{\left(u,v\right)}}}^{\Omega}\left(I\right) & = & \frac{w_{x}^{'}\left(u,v\right)-L_{min_{{x}}}}{\delta}\\
m_{x_{{\left(u,v\right)}}}^{\Omega}\left(P\right) & = & \frac{L_{max_{{x}}}-w_{x}^{'}\left(u,v\right)}{\delta}\\
m_{x_{{\left(u,v\right)}}}^{\Omega}\left(\left\{ I,P\right\} \right) & = & 1-\left(m_{{x_{{\left(u,v\right)}}}}^{\Omega}\left(I\right)
+m_{{x_{\left(u,v\right)}}}^{\Omega}\left(P\right)\right)
\end{eqnarray}
where $L_{min_{{x}}}=\min_{\left(a,b\right)\in E}w_{x}^{'}\left(a,b\right)$,
$L_{max_{{x}}}=\max_{\left(a,b\right)\in E}w_{x}^{'}\left(a,b\right)$ and\\
 $\delta=L_{max_{{x}}}-L_{min_{{x}}}+\beta$, $\beta\in\left[0,1\right]$ is used
to model an imprecise knowledge adding belief on ignorance, \textit{i.e.}
the set $\left\{ I,P\right\} $, to model our uncertainty. Consequently,
we obtain three BBA distributions defined on $\Omega$, \textit{i.e.}
follow BBA, mention BBA and retweet BBA, for each link in the network.
We combine them using the Dempster's rule of combination (equation
(\ref{eq:dempster rule})). As a result, we get a mass distribution,
$m_{{\left(u,v\right)}}^{\Omega}$, for each link. {The novelty of this BBA is that it fuses} many influence aspects:
\begin{enumerate}
\item The strength of the link between $u$ and $v$ in the network structure
that is measured by the mean of the follow weight.
\item Information exchange and propagation activities between users that
is considered through the mention and the retweet weights respectively.
\item The fact of being more influencer if you are connected to influencer users.
\end{enumerate}
Finally, the influence of $u$ on $v$ is defined as the amount of
mass given to $\left\{ I\right\} $ as:

\begin{equation}
Inf\left(u,v\right)=m_{{\left(u,v\right)}}^{\Omega}\left(I\right)
\end{equation}

Next, we define the amount of influence given to a set of nodes $S\subseteq V$
for influencing a user $v\in V$. We present two estimation ways;
in the first one, we consider the influence on the directly connected
nodes to $S$ and in the second one, we consider also nodes that are
connected to neighbors of $S$. The work of Chen \textit{et al.} \citep{Chen2010}
justifies our formulas. They {affirm} that when the product have some
quality issues, it is more adaptable to choose influencers that have
many immediate neighbors. In fact, when the influence propagates in
many hops in the network, it may fall on a user that dislikes the
product. Besides, when the product have a high quality, we can choose
users that have a large reachable set. We estimate the influence of
$S$ on a user $v$ as follows:

\begin{eqnarray}
Inf\left(S,v\right) & = & \begin{cases}
1 & \textrm{if}\, v\in S\\
{\displaystyle \sum_{u\in S}Inf\left(u,v\right)} & \textrm{Otherwise}
\end{cases}\label{eq:level1}\\
Inf\left(S,v\right) & = & \begin{cases}
1 & \textrm{if}\, v\in S\\
{\displaystyle \sum_{u\in S}{\displaystyle }\sum_{x\in D_{IN}\left(v\right)\cup\left\{ v\right\} }Inf\left(u,x\right).Inf\left(x,v\right)} & \textrm{Otherwise}
\end{cases}\label{eq:level2}
\end{eqnarray}
such that $Inf\left(v,v\right)=1$ and $D_{IN}\left(v\right)$ is
the set of nodes in the indegree of $v$. 
\\
Finally, we define the influence spread $\sigma_{Bel}\left(S\right)$
under the evidential model as the total influence given to $S\subseteq V$
from all nodes in the social network:

\begin{equation}
\sigma_{Bel}\left(S\right)=\sum_{v\in V}Inf\left(S,v\right)
\end{equation}

In the spirit of the IM problem, as defined by \citep{Kempe03}, $\sigma_{Bel}\left(S\right)$
is the objective function to be maximized.

\subsection{Evidential influence maximization}

In this section, we present the evidential influence maximization
model. Its purpose is to find a set of nodes $S$ that maximizes the
objective function $\sigma_{Bel}\left(S\right)$. Given a directed
social network $G=\left(V,\, E\right)$, an integer $k\leq|V|$, a
tweet table, $T$, that contains user's tweets that are published
in a period of time $t$ ($t$ is a week for example) and an activity
table, $A$, that contains mentions and retweets that are made in
$t$. The goal is to find a set of users $S\subseteq V,$ $|S|=k$,
that maximizes $\sigma_{Bel}\left(S\right)$.

\begin{thm} $\sigma_{Bel}\left(S\right)$ is monotone and sub-modular.\end{thm}

\begin{pf} $\sigma_{Bel}\left(S\right)$ is monotone, \textit{i.e.} $\sigma_{Bel}\left(S\right)\leq\sigma_{Bel}\left(T\right),\, S\subseteq T$.
In fact, $\sum_{v\in V}Inf\left(S,v\right)\leq\sum_{v\in V}Inf\left(T,v\right)$.
$\sigma_{Bel}\left(S\right)$ is sub-modular if and only if $\sigma_{Bel}\left(S\cup\left\{ x\right\} \right)-\sigma_{Bel}\left(S\right)\geq\sigma_{Bel}\left(T\cup\left\{ x\right\} \right)-\sigma_{Bel}\left(T\right),\, S\subseteq T$,
\textit{i.e.} the marginal gain of $x$ with respect to $T$ is no
more than the marginal gain of $x$ with respect to $S$. In the case
were $x\in S$, we have\\
 $\sigma_{Bel}\left(S\cup\left\{ x\right\} \right)-\sigma_{Bel}\left(S\right)=\sigma_{Bel}\left(T\cup\left\{ x\right\} \right)-\sigma_{Bel}\left(T\right)=0,\, S\subseteq T$.
If $x\notin S$ we have two alternatives; if we use the formula (\ref{eq:level1}),
we proven that 
\begin{eqnarray}
MG_{S}\left(x\right) & = & 1+\sum_{v\in V\setminus S}Inf\left(x,v\right)\label{eq:mgL1}\\
MG_{T}\left(x\right) & = & 1+\sum_{v\in V\setminus T}Inf\left(x,v\right)
\end{eqnarray}
 Where $MG_{S}\left(X\right)=\sigma_{Bel}\left(S\cup\left\{ x\right\} \right)-\sigma_{Bel}\left(S\right)$
and $MG_{T}\left(x\right)=\sigma_{Bel}\left(T\cup\left\{ x\right\} \right)-\sigma_{Bel}\left(T\right)$.
In the second case, \textit{i.e.} the case of the formula (\ref{eq:level2}),
we have 
\begin{eqnarray}
MG_{S}\left(x\right) & = & 1+\sum_{v\in V\setminus S\,}\sum_{a\in D_{IN}\left(v\right)\cup\left\{ v\right\} }Inf\left(x,a\right).Inf\left(a,v\right)\label{eq:mgL2}\\
MG_{T}\left(x\right) & = & 1+\sum_{v\in V\setminus T\,}\sum_{a\in D_{IN}\left(v\right)\cup\left\{ v\right\} }Inf\left(x,a\right).Inf\left(a,v\right)
\end{eqnarray}
 In the two cases we have $S\subseteq T$ then $|V\setminus S|\geq|V\setminus T|$
which proves the sub-modularity of $\sigma_{Bel}\left(S\right)$.
~~~$\square$
\end{pf}

\begin{thm} Influence maximization under the evidential model is NP-Hard.\end{thm} 

\begin{pf} 
To demonstrate the hardness of our approach, we show that the function given by the equation (\ref{eq:level2}) can be seen as a particular case of the function of the CD model \citep{Goyal12} that was shown to be NP Hard. If we suppose that we have one action $a$ then 
\begin{eqnarray}
\gamma\left(u,v\right)\left(a\right)=\Gamma\left(u,v\right)\left(a\right)=Inf\left(u,v\right)
\end{eqnarray}
 and 
\begin{eqnarray}
\Gamma\left(S,v\right)=\begin{cases}
1 & \textrm{if}\, v\in S\\
{\displaystyle \sum_{x\in D_{IN}\left(v\right)}Inf\left(S,x\right).Inf\left(x,v\right)} & \textrm{Otherwise}
\end{cases}
\end{eqnarray}

$Inf\left(S,v\right)$ can be seen as $\Gamma\left(S,v\right)$ of the CD model by considering only two hops between neighbors while estimating influence. Then we prove that ``2 Levels'' model is NP Hard. Also, we can write the function given by (\ref{eq:level1}) of the ``1 Level'' model by a sum of the product of two functions. Then, we show that the ``1 Level'' model is NP Hard. ~~~ $\square$
\end{pf}

We showed that the influence maximization under the evidential model is NP-Hard, besides, the influence spread function is monotone and sub-modular. Therefore, the greedy algorithm performs good approximation for the optimal solution especially when we use it with formula (\ref{eq:mgL1}) or formula (\ref{eq:mgL2}) that computes the marginal gain of a candidate node $x$. We choose the cost effective lazy-forward algorithm (CELF) \citep{Leskovec07b} which is a two pass modified greedy algorithm that is proved to be about 700 time faster than the basic greedy algorithm. CELF exploits the sub-modularity property of the function to be maximized, in fact, sub-modularity guarantees that marginal benefits decrease with the solution size, hence, instead of computing the marginal benefit of each expected node at each iteration, CELF computes it in the first iteration and keeps an ordered list of nodes according to their marginal benefits value for the next iteration. In the next iteration, it re-evaluates the marginal benefit for the top node then it resorts the node list. If the top node {maintains} its position, it will be chosen elsewhere  {CELF re-evaluates} the marginal benefit for the new top node and so on. Algorithm (\ref{alg:CELF-based-evidential}) shows steps of the CELF based evidential influence maximization algorithm.

\begin{algorithm}
\caption{CELF based evidential influence maximization algorithm\label{alg:CELF-based-evidential}}
\Begin{

 \emph{$S=\emptyset$}\;\tcp{$S$: the set of seed nodes}
 \emph{$Q=\emptyset$}\;\tcp{$Q$: sorted list in decreasing order according to the marginal gain of nodes}
\For{each $node\in V$}{
\emph{$marginalGain(node)$}\; \tcp{$marginalGain()$ estimate the marginal gain of the node}
 \emph{$Q.add(node)$}\;}
 
\emph{$nodeMax\leftarrow Q.pop()$}\;
\emph{$S.add(nodeMax)$}\;
\While {$\mid S\mid\leq k$}{

\emph{$nodeMax\leftarrow Q.pop()$}\;
\emph{$updateMarginalGain(nodeMax)$}\; \tcp{We use formula \ref{eq:mgL1} or \ref{eq:mgL2} to update the marginal gain}
\lIf{$nodeMax.MG\geq Q.getFirst().MG$} {$S.add(nodeMax)$}
\lElse{$Q.add(nodeMax)$}}

}

\end{algorithm}

\section{Experiments and results}
\begin{sidewaysfigure}
\includegraphics[scale=0.45]{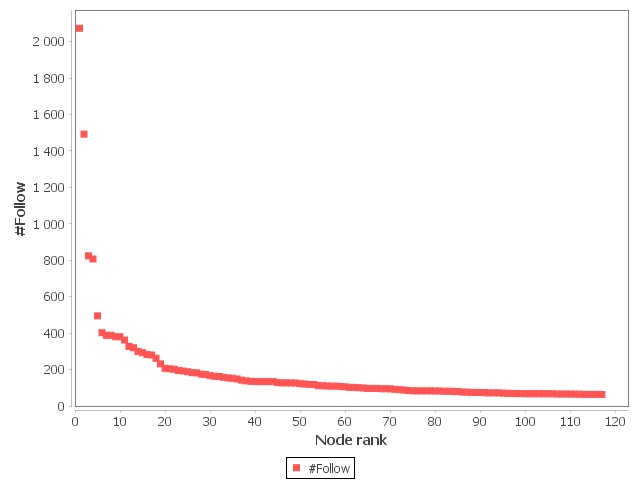} \includegraphics[scale=0.45]{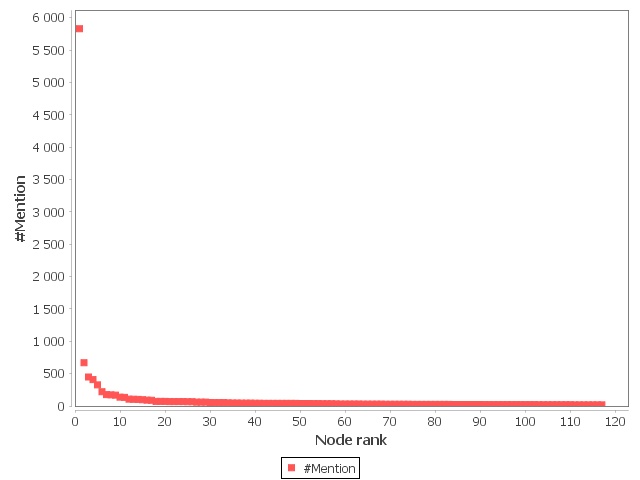}

\includegraphics[scale=0.45]{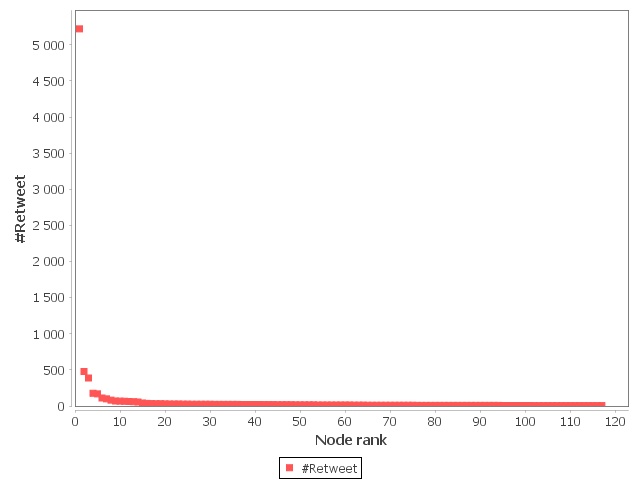} \includegraphics[scale=0.45]{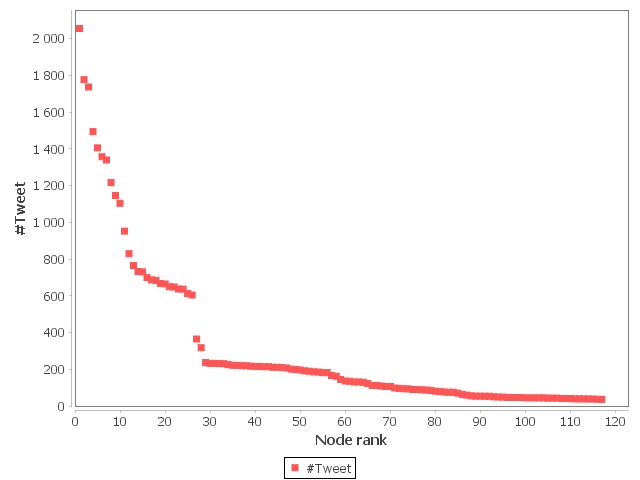}

\caption{Data distributions\label{fig:Data-distributions} }
\end{sidewaysfigure}

In this section, we conduct some experiments on real world data to compare the proposed models with existing ones. We used the library Twitter4j%
\footnote{Twitter4j is a java library for the Twitter API, it is an open-sourced
software and free of charge and it was created by Yusuke Yamamoto.
More details can be found in http://twitter4j.org/en/index.html.%
} which is a java implementation of the Twitter API to collect Twitter
data. We crawled the Twitter network for the period between 08/09/2014
and 03/11/2014 and we filtered our data by keeping only tweets that
talk about smartphones and users that have at least one tweet in the 
data base. Table (\ref{tab:Statistics}) shows some statistics about
the collected data and Figure (\ref{fig:Data-distributions}) displays
{users'} ranks based on follow, mention, retweet and tweet across our
data.

\begin{table}
\caption{Statistics of the data set\label{tab:Statistics}}
\begin{centering}
\begin{tabular}{|c|c|c|c|c|}
\hline 
\#User & \#Tweet & \#Follow & \#Retweet & \#Mention\tabularnewline
\hline 
36274 & 251329 & 71027 & 9789 & 20300\tabularnewline
\hline 
\end{tabular}
\par\end{centering}

\end{table}

To study the accuracy of the proposed influence maximization models, we use a generated dataset. In fact, we generated data in such a way one can know the influencers. Then, we obtain a useful dataset to study the accuracy of the proposed influence maximization models. Social network structure has some special characteristics that differentiate it from ordinary graphs like the small world assumption \citep{Newman10}. For this reason, we chose to use a real world structure. Then, we selected a random sampling of the collected network from Twitter. The sampled network contains 1010 nodes and 6906 directed links between them. In a second step, we selected a set of users that have at least 15 outlinks. As a result, we have got a set of 108 users. Next, we define, randomly, the influence on each link in the network and the selected 108 users are defined as influencers by setting maximum influence values in their outlinks. The minimum value of influence given to an influencer is a parameter {of} the random process.


\subsection{Experiments configuration}

In our experiments, we compare the proposed evidential influence maximization
model with:
\begin{itemize}
\item Credit distribution (CD) model that we find the closest in its principle
to our model.
\item Independent cascade model with uniform edge probabilities (UN ICM)
equals to $1\%$.
\item ICM with trivalency edge probabilities (TV ICM), \textit{i.e.} chosen
randomly from $\left\{ 10\%,\,1\%,\,0.1\%\right\} $.
\item Weighted cascade (WC ICM) \textit{i.e.} ICM with edge probability
of $\left(u,v\right)$ equals to $\frac{1}{D_{u}}$.
\item Linear threshold model (LTM) with uniform edge weights $\omega\left(u,v\right)=1\%$
and random threshold $\theta_{u}$ for each node.
\end{itemize}
To fix ICM edge probabilities and LTM weights we followed the experiments
of previous works \citep{Kempe03,Goyal12} and we run the algorithms
10000 times with the Monte-Carlo simulation. Furthermore, to examine
the quality of the selected seeds by each method we fixed four comparison
criteria which are: the number of followers, \textit{\#Follow}, the
number of tweets, \textit{\#Tweet}, the number of times the user was
mentioned and retweeted, \textit{\#Mention} and \textit{\#Retweet}.
In fact, we assume that if a user is an influencer on Twitter he would 
be necessarily: very active {and} he has a lot of tweets, he is followed
by many users in the network that are interested by his news, he is
frequently mentioned in {others'} tweets and his tweets are retweeted
several times.

\subsection{Results and discussion}

The main goal of our experiments is to show the performance of the
proposed approach. We denote by ``1 Level'' the evidential influence
maximization model that uses the formula (\ref{eq:level1}) and by
``2 Levels'' the evidential model with the formula (\ref{eq:level2}). 

\begin{sidewaysfigure}
\begin{centering}
\subfloat[Accumulated follow\label{fig:Cumulated-follow}]{\includegraphics[scale=0.55]{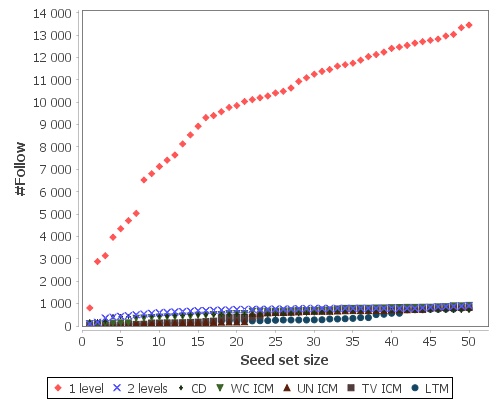}

} \subfloat[Accumulated mention\label{fig:Cumulated-mention}]{\includegraphics[scale=0.55]{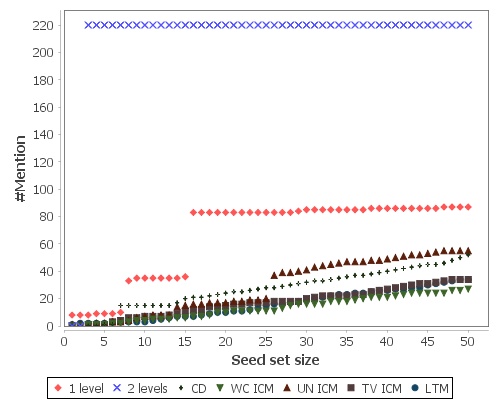}

}
\par\end{centering}

\begin{centering}
\subfloat[Accumulated retweet\label{fig:Cumulated-retweet}]{\includegraphics[scale=0.55]{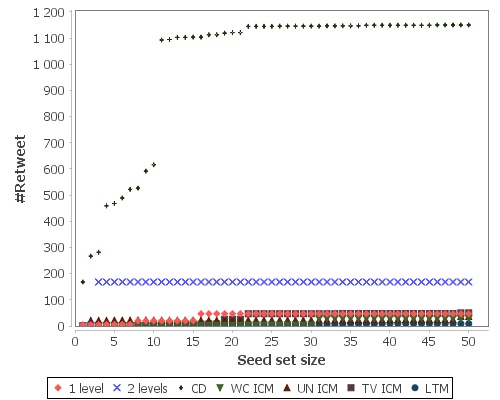}

} \subfloat[Accumulated tweet\label{fig:Cumulated-tweet}]{\includegraphics[scale=0.55]{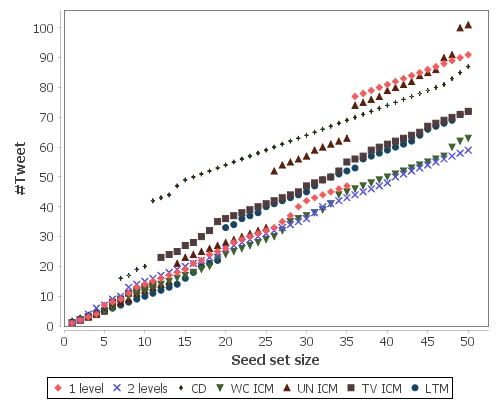}

}
\par\end{centering}

\caption{Comparison between the proposed models (``1 Level'' and ``2 Levels''), CD model, ICM and LTM\label{fig:ComparisionLTMICM}}
\end{sidewaysfigure}

In Figure (\ref{fig:ComparisionLTMICM}), we compare the proposed
approach to some existing ones \textit{i.e.} CD, ICM and LTM. As it
was very hard to turn the basic models on the {whole} dataset, this fact
was shown by previous works like \citep{Goyal12}, we used a sampling
of 1010 nodes from the original data. In Figure (\ref{fig:Cumulated-follow}),
we observe that ``2 Levels'', LTM, UN ICM, TV ICM and CD detect weakly connected
users at first. However, we observe that the {``1 Level''} model of
the proposed approach detects strongly connected users. Figure (\ref{fig:Cumulated-mention})
shows that most scatter plots are close to each {other} except
that of { ``1 Level'' and} ``2 Levels'' that detected highly mentioned users. In Figure (\ref{fig:Cumulated-retweet}), we observe that the best results are given by the CD model. Besides, the ``2 Levels'' has successfully detect highly retweeted users, also, we see that ``1 Level'', WC ICM, UN ICM and LTM have almost
close scatter plots. Finally, Figure (\ref{fig:Cumulated-tweet})
shows that ``1 Level'' model, ``2 Levels'' model, WC ICM, UN ICM,
TV ICM, LTM {and CD} detected active users.

 From these {observations}, we conclude that ``1 Level'' and ``2 Levels'' models of
the proposed approach detected influencer users that are active and
have a good position in the network {allowing} them to propagate their
messages in a short time. Also, we conclude that {``1 Level''} is the best model in selecting influencer users. In fact, it chooses users that have a good compromise between the four criteria, \textit{i.e.} \#Follow, \#Mention, \#Retweet and \#Tweet. In Table (\ref{tab:Running-time}),
we present the running time in milliseconds of methods in the
experiment of Figure (\ref{fig:ComparisionLTMICM}). {In fact, all the experimented models are proven to be NP-Hard \citep{Kempe03, Goyal12}.} As shown in Table (\ref{tab:Running-time}) the proposed models are faster than existing models. In fact, the ``1 Level'' model {gives} its results in {62} milliseconds and the ``2 Levels'' in {869} milliseconds while LTM and ICM needs many thousands of seconds to {give} their results.

\begin{table}
\caption{Running time in milliseconds\label{tab:Running-time}}
\begin{centering}
\begin{tabular}{|c|c||c|c|}
\hline 
\textbf{Model} & \textbf{Time} & \textbf{Model} & \textbf{Time}\tabularnewline
\hline 
1 Level & \textbf{{62}} & TV ICM & 7267904\tabularnewline
\hline 
2 Level & \textbf{{869}} & UN ICM & 4844867\tabularnewline
\hline 
CD & 4654 & WC ICM & 4295455\tabularnewline
\hline 
LTM & 65963285 &  & \tabularnewline
\hline 
\end{tabular}
\par\end{centering}

\end{table}

\begin{sidewaysfigure}
\begin{centering}
\subfloat[Accumulated follow Comparison\label{fig:Cumulated-follow-Comparision3000}]{\includegraphics[scale=0.55]{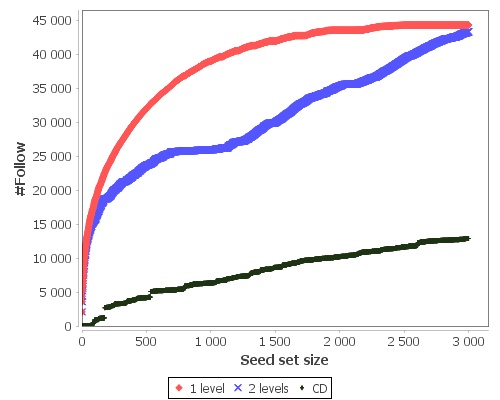}

} \subfloat[Accumulated mention comparison\label{fig:Cumulated-mention-comparision3000}]{\includegraphics[scale=0.55]{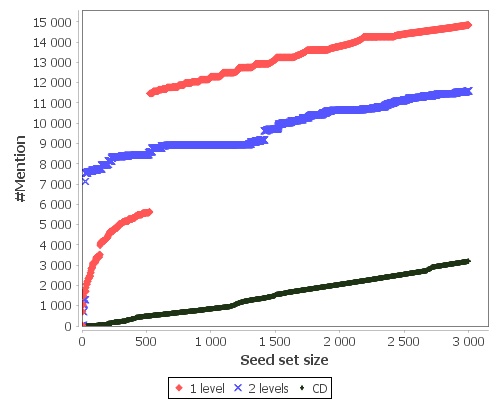}

}
\par\end{centering}

\begin{centering}
\subfloat[Accumulated retweet comparison\label{fig:Cumulated-retweet-comparision3000}]{\includegraphics[scale=0.55]{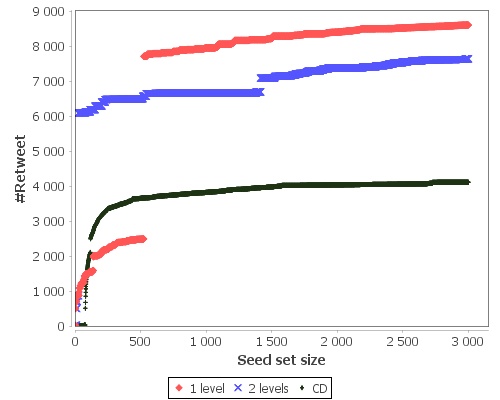}

} \subfloat[Accumulated tweet comparison\label{fig:Cumulated-tweet-comparision3000}]{\includegraphics[scale=0.55]{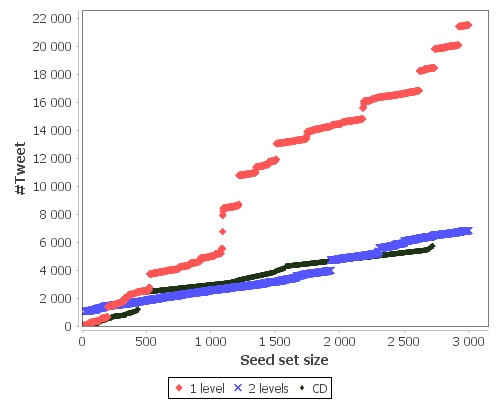}

}
\par\end{centering}

\caption{Comparison between the proposed models (1 level and 2 levels) and
credit distribution (CD) model with S size = 3000\label{fig:ComparisionWithCD3000}}
\end{sidewaysfigure}

\begin{sidewaysfigure}
\begin{centering}
\subfloat[Accumulated follow Comparison\label{fig:Cumulated-follow-Comparision-1}]{\includegraphics[scale=0.55]{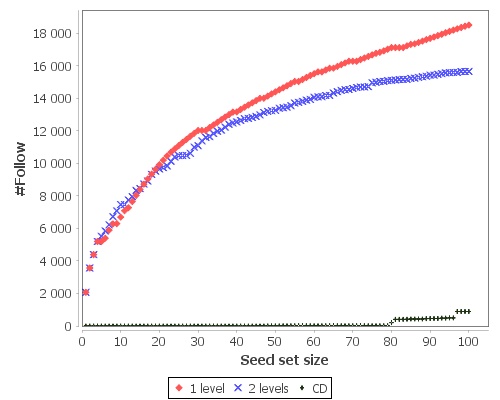}

} \subfloat[Accumulated mention comparison\label{fig:Cumulated-mention-comparision-1}]{\includegraphics[scale=0.55]{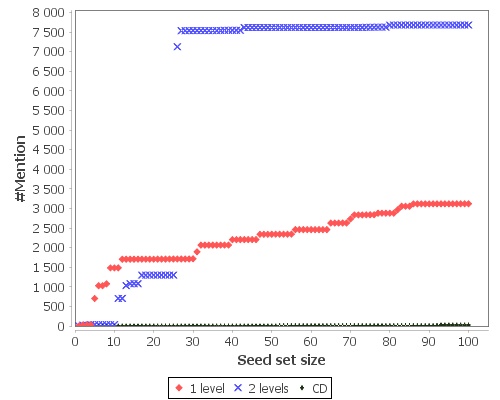}

}
\par\end{centering}

\begin{centering}
\subfloat[Accumulated retweet comparison\label{fig:Cumulated-retweet-comparision-1}]{\includegraphics[scale=0.55]{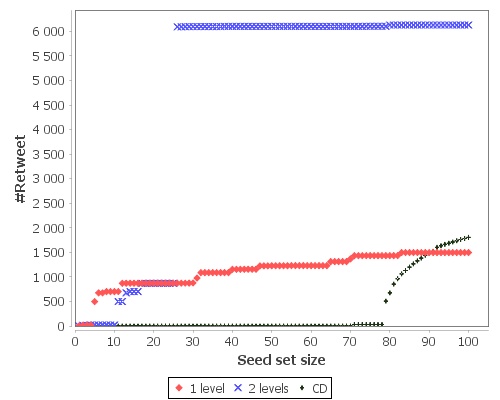}

} \subfloat[Accumulated tweet comparison\label{fig:Cumulated-tweet-comparision-1}]{\includegraphics[scale=0.55]{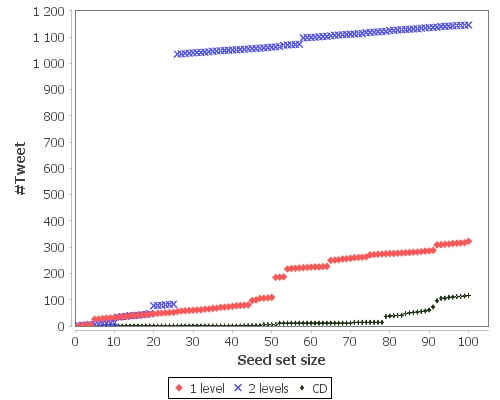}

}
\par\end{centering}

\caption{Comparison between the proposed models (1 level and 2 levels) and
credit distribution (CD) model with S size = 100\label{fig:ComparisionWithCD-1}}
\end{sidewaysfigure}

As credit distribution model is the closest in its principle to the proposed models, we use the {whole} dataset to compare it with ``1 Level'' and ``2 Levels'' according to the accumulated \#Follow (Figures (\ref{fig:Cumulated-follow-Comparision3000})
and (\ref{fig:Cumulated-follow-Comparision-1})), the accumulated
\#Mention (Figures (\ref{fig:Cumulated-mention-comparision3000})
and (\ref{fig:Cumulated-mention-comparision-1})), the accumulated
\#Retweet (Figures (\ref{fig:Cumulated-retweet-comparision3000})
and (\ref{fig:Cumulated-retweet-comparision-1})) and the accumulated
\#Tweet (Figures (\ref{fig:Cumulated-tweet-comparision3000}) and
(\ref{fig:Cumulated-tweet-comparision-1})) of seed set nodes.

Figures (\ref{fig:ComparisionWithCD3000}) and (\ref{fig:ComparisionWithCD-1})
show the performance of the proposed models (1 Level and 2 Levels)
against the credit distribution (CD) model. In fact we see that the evidential
influence maximization approach {detects} influencer spreaders that have
a good compromise between \#Follow, \#Mention, \#Retweet and \#Tweet.
We observe that they detected seeds that are followed by many users. Indeed, 
in Figure (\ref{fig:Cumulated-follow-Comparision-1}) we see that
the first 10 seeds are followed by over 6000 users while there {are}
no followers for the first 10 seeds that are detected by CD model.
According to Figures (\ref{fig:Cumulated-mention-comparision3000})
and (\ref{fig:Cumulated-mention-comparision-1}), detected seeds with
the ``1 Level'' and the ``2 Levels'' models are mentioned many
times whereas the CD model starts to detect mentioned users after
over 93 seed {nodes} detected. In Figure (\ref{fig:Cumulated-retweet-comparision3000})
we see that the CD model has successfully detected users that were
retweeted a lot. However, Figure (\ref{fig:Cumulated-retweet-comparision-1})
shows that this model started to detect retweeted users only after
about 70 seed nodes detected while the evidential influence maximization
models start detecting them from the second seed. Finally, Figures
(\ref{fig:Cumulated-tweet-comparision3000}) and (\ref{fig:Cumulated-tweet-comparision-1})
show the accumulated activity size of the detected seeds that is measured
by their number of tweets. We see that the CD model has the same behavior
as in the retweet scatter plot and it starts to detect active users
after about 50 seeds, in the other hand, the proposed approach {demonstrates}
its performance in detecting active users from the second seed detected.
From Figures (\ref{fig:ComparisionWithCD3000}) and (\ref{fig:ComparisionWithCD-1})
we conclude that the proposed evidential models are better than the
CD model in that the evidential models provide a good compromise between
the four influence criteria (\#Follow, \#Mention, \#Retweet and
\#Tweet) in Twitter. In fact, the selected influencer spreaders are active, have a
good position in the network, also, they are highly mentioned in
others tweets and their tweets are highly retweeted. {However, the CD model starts selecting followed user's after about 40 seeds, mentioned users after about 93 seeds, retweeted users after about 70 seeds and active users after about 50 seeds.}

\begin{figure}
\begin{centering}
\includegraphics[scale=0.5]{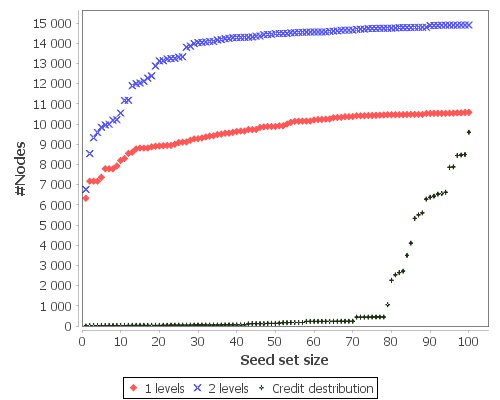}
\par\end{centering}

\caption{The {dependence} of the number of affected nodes to the size of S\label{fig:NbrNodesVsS}}

\end{figure}

In Figure (\ref{fig:NbrNodesVsS}), we examine the number of distinct
affected nodes that are connected to the influencers and to their
neighbors. We observe that CD model detected about 40 isolated
users at first and it started to detect users that are followed by many other users from the seed node 80.
 In the other hand, we notice a different behavior of scatter plots of  ``1 Level'' and
 ``2 Levels'' models. In fact, in Figure (\ref{fig:Cumulated-follow-Comparision3000}) ``1 Level'' scatter plot is upper than the scatter plot of ``2 Levels'' model.
However, in Figure (\ref{fig:NbrNodesVsS}) we observe that ``2
Levels'' scatter plot is upper than the scatter plot
of ``1 Level'' model. From these observations, we conclude that the
``2 Levels'' model detects influencer spreader that are connected
to highly followed users and the ``1 Level'' model detects highly
followed influencer spreaders. Also, we conclude that our models are better in detecting seeds than CD model. Indeed, ``1 Level'' and ``2 Levels'' models detect highly connected seeds at first. However, the CD model selects about 40 isolated seeds before starting to detect some followed seeds.

\begin{sidewaysfigure}
\begin{centering}
\includegraphics[scale=0.6]{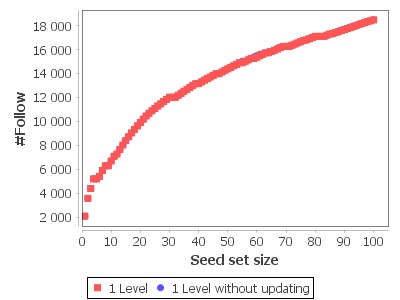}
\includegraphics[scale=0.6]{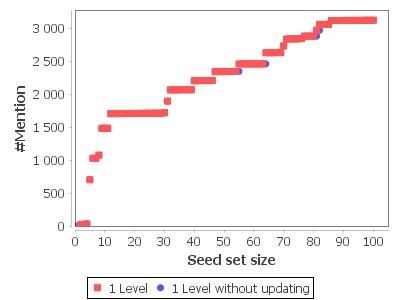}
\par\end{centering}

\begin{centering}
\includegraphics[scale=0.6]{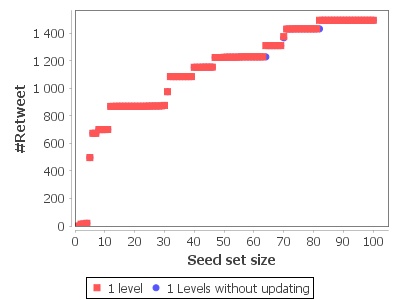}
\includegraphics[scale=0.6]{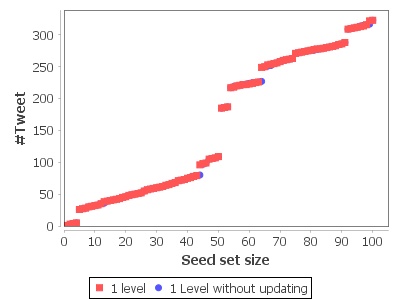}
\par\end{centering}

\caption{Impact of the weight updating step on influence maximization results: 1 Level\label{fig:Impact-1}}
\end{sidewaysfigure}

\begin{sidewaysfigure}
\begin{centering}
\includegraphics[scale=0.5]{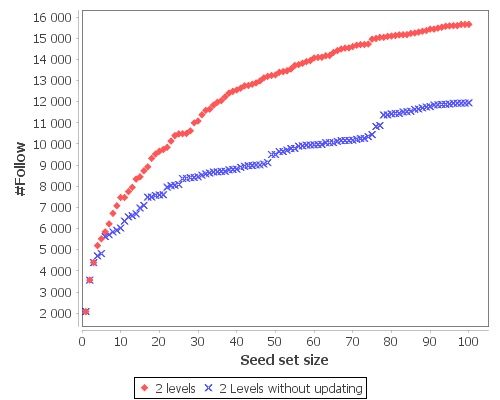}
\includegraphics[scale=0.5]{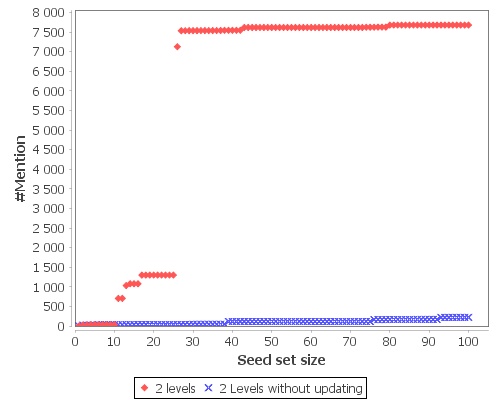}
\par\end{centering}

\begin{centering}
\includegraphics[scale=0.5]{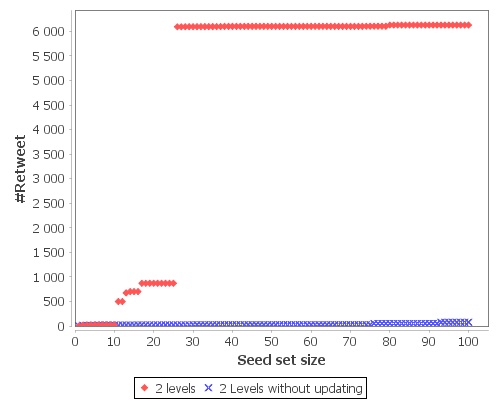}
\includegraphics[scale=0.5]{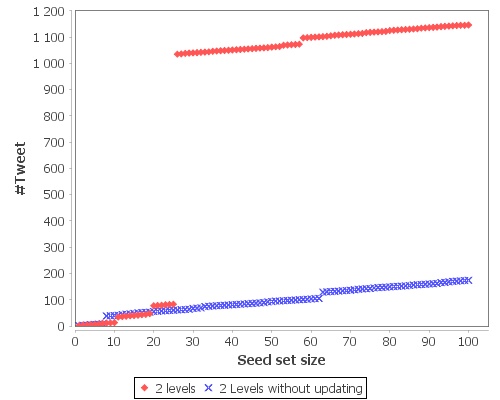}
\par\end{centering}

\caption{Impact of the weight updating step on influence maximization results: 2 Levels\label{fig:Impact-2}}
\end{sidewaysfigure}

Our goal in the experiments of Figures (\ref{fig:Impact-1}) and (\ref{fig:Impact-2})
is to show the impact of considering the fact of \textit{``being more influencer
if you are connected to influencer users''} on the influence maximization
results. This fact is considered in the second step, \textit{i.e.}
``Updating step'', of our influence estimation process. Then we
compare  ``1 Level'' (Figure (\ref{fig:Impact-1})) and 
``2 Levels'' (Figure (\ref{fig:Impact-2})) models with and without
the updating step. Figure (\ref{fig:Impact-1}) shows that the difference
in the ``1 Level'' is not very significant. However, in Figure (\ref{fig:Impact-2})
we see that the updating step ameliorates the influence maximization
results for the ``2 Levels'' model. Indeed, when we consider the assumption of ``being more influencer
if you are connected to influencer users'', the ``2 Levels'' model detects better seeds than the model without considering this assumption.

\begin{figure}
	\centering
		\includegraphics[width=0.750\textwidth]{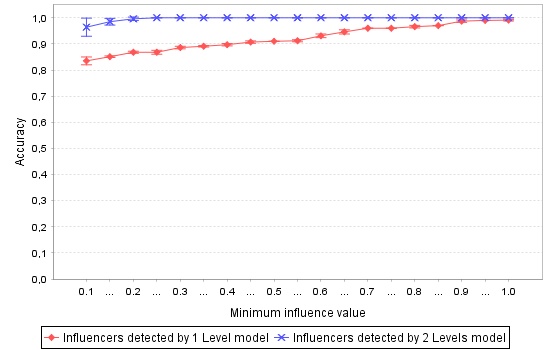}
	\caption{Accuracy of the proposed influence maximization models on generated data}
	\label{fig:AccuracyComparision}
\end{figure}

We introduce a last experiment to study and compare the accuracy of the proposed influence maximization models. For this purpose we run the ``1 Level'' and the ``2 Levels'' models on the generated data set in which the set of influencers is known. Then, we compute the hit ratio, \textit{i.e.} the percent of correctly detected influencers, in order to compare the set of predicted $k$ influencers with the known set of influencers. As we did this experiment on generated data, then we varied the minimum value of influence given to an influencer. We fixed the size of the seed set $k$ to 50 and we repeated the random process ten times. We obtained the results shown in Figure \ref{fig:AccuracyComparision}.

According to  Figure \ref{fig:AccuracyComparision}, the proposed models have a good accuracy in detecting influencers. This figure shows the performance of the proposed models. In fact, even with a small influence value, 0.1, the experimented models succeed in detecting influencers with a good accuracy that is no less than 
$83\% \pm 0.01$. Besides, we notice that the ``2 Levels'' model  have the highest accuracy values. 

\section{Conclusion}

In this work, we introduced two new evidential influence maximization models. The proposed two models are based on a new influence estimation measure for Twitter that considers many influence aspects like the importance of the user in the network structure and the popularity of user's tweets messages. Then, we used the CELF algorithm to solve the influence maximization problem under the proposed evidential approach. To show the performance of the proposed models, we conducted some experiments to compare it with existing models. We proved that the proposed models are better than existing ones in selecting influencer users for the Twitter social network. In fact, the selected seeds performs a good compromise between the basic criteria (\#Follow, \#Mention, \#Retweet and \#Tweet) of influence in Twitter. However, we find that the CD model, for example, fails to detect good influencers at first. In fact, it detects about 40 isolated users before starting to detect followed ones. This is not the case of the proposed models. Indeed, the first selected user by our models has about 2000 followers.

In future works, we will search to improve the proposed influence measure by considering the user's profile, the topic of the message and more levels of influence in the network. {Another} important objective is to adapt the proposed influence maximization model to other social networks like Facebook and LinkedIn. Finally, we will search to test the evidential influence maximization approach with larger data bases.

\section{Acknowledgment}

This work was done within the MOBIDOC device launched under the Support Project to the Research and Innovation System (PASRI), funded by the European Union and managed by the National Agency for the Promotion of Scientific Research (ANPR). Also, we thank the \textquotedbl{}Centre
d\textquoteright{}Etude et de Recherche des Télécommunications\textquotedbl{}
(CERT) for their support.

\section{References}
\bibliographystyle{IEEEtran}
\bibliography{biblio}

\end{document}